\documentclass{emulateapj}

\usepackage{xspace}
\usepackage{amssymb}
\usepackage{graphicx}
\usepackage{booktabs}
\usepackage{hyperref}
\usepackage{txfonts}
\usepackage{hyperref}
\usepackage{graphicx, float}
\usepackage{gensymb}
\usepackage{natbib}
\usepackage{epsfig, natbib}

\shorttitle{The metal-poor knee in the Fornax Dwarf Spheroidal Galaxy}
\shortauthors{Hendricks et al.}

\begin{document}

\title
{The metal-poor knee in the Fornax Dwarf Spheroidal Galaxy\altaffilmark{$\dagger$}}
\author{Benjamin Hendricks\altaffilmark{1}, Andreas Koch\altaffilmark{1}, Gustavo A. Lanfranchi\altaffilmark{2}, Corrado Boeche\altaffilmark{3}, Matthew Walker\altaffilmark{4}, Christian I. Johnson\altaffilmark{5,}\altaffilmark{6}, Jorge Pe\~{n}arrubia\altaffilmark{7}, Gerard Gilmore\altaffilmark{8}}
\altaffiltext{$\dagger$}{This article is based on observations made with ESO
Telescopes at the Paranal Observatory under programme 082.B-0940(A).}

\affil{$^1$ Zentrum f\"ur Astronomie der Universit\"at Heidelberg, Landessternwarte, K\"onigstuhl 12, 69117, Heidelberg, Germany; ben.hendricks@lsw.uni-heidelberg.de}
\affil{$^2$ N\'ucleo de Astrof\'\i sica Te\'orica, Universidade
Cruzeiro do Sul, R. Galv\~ao Bueno 868, Liberdade, 01506-000, S\~ao Paulo, SP, Brazil}
\affil{$^3$ Zentrum f\"ur Astronomie der Universit\"at Heidelberg, Astronomisches Rechen-Institut, M\"onchhofstr. 12-14, 69120, Heidelberg, Germany}
\affil{$^4$McWilliams Center for Cosmology, Carnegie Mellon University, 5000 Forbes Ave., Pittsburgh, PA 15213, USA}
\affil{$^5$Harvard-Smithsonian Center for Astrophysics, 60 Garden Street, MS-15, Cambridge, MA 02138, USA}
\affil{$^6$ Clay Fellow}
\affil{$^7$ Institute for Astronomy, University of Edinburgh, Royal Observatory, Blackford Hill, Edinburgh EH9 3HJ, UK}
\affil{$^8$ Institute of Astronomy, Cambridge University, Madingley Rd, Cambridge CB3 OHA, UK}

\begin{abstract}
We present $\mathrm{\alpha}$-element abundances of Mg, Si, and Ti for a large sample of field stars in two outer fields of the Fornax dwarf spheroidal galaxy (dSph), obtained with VLT/GIRAFFE ($R\sim16,000$). Due to the large fraction of metal-poor stars in our sample, we are able to follow the $\mathrm{\alpha}$-element evolution from $\mathrm{[Fe/H]}\approx-2.5$ continuously to $\mathrm{[Fe/H]}\approx-0.7$. For the first time we are able to resolve the turnover from the Type II supernovae (SNe) dominated, $\mathrm{\alpha}$-enhanced plateau down to subsolar $\mathrm{[\alpha/Fe]}$ values due to the onset of SNe Ia, and thus to trace the chemical enrichment efficiency of the galaxy. Our data support the general concept of an $\mathrm{\alpha}$-enhanced plateau at early epochs, followed by a well-defined ``knee'', caused by the onset of SNe Ia, and finally a second plateau with sub-solar $\mathrm{[\alpha/Fe]}$ values. We find the position of this knee to be at $\mathrm{[Fe/H]}\approx-1.9$ and therefore significantly more metal-poor than expected from comparison with other dSphs and standard evolutionary models. Surprisingly, this value is rather comparable to the knee in Sculptor, a dSph $\sim10$ times less luminous than Fornax. Using chemical evolution models, we find that both the position of the knee as well as the subsequent plateau at sub-solar level can hardly be explained unless the galaxy experienced several discrete star formation events with a drastic variation in star formation efficiency, while a uniform star formation can be ruled out. One possible evolutionary scenario is that Fornax experienced one or several major accretion events from gas-rich systems in the past, so that its current stellar mass is not indicative of the chemical evolution environment at ancient times. If Fornax is the product of several smaller buildings blocks, this may also have implications of the understanding on the formation process of dSphs in general. 
\end{abstract}

\section{Introduction}
\label{chap_01}

Dwarf spheroidal (dSph) galaxies in the Local Group are the smallest, closest and most abundant galaxies in the Universe and are therefore excellent laboratories to study galactic star formation history (SFH) and chemical evolution on the smallest scale.
Among these, Fornax is one of the most massive (stellar mass $\sim10^7 M_{\odot}$, \citealt{de_Boer_12}) and most metal-rich ($\langle\mathrm{[Fe/H]}\rangle \approx-1.0 $) dSphs at a distance of $147\,\mathrm{kpc}$ (\citealt{McConnachie_12}), and one of only two systems known to host a globular cluster (GC) population (the other one is Sagittarius). 
Fornax displays complex chemical and dynamical properties: the metallicity-distribution-function (MDF) from several spectroscopic studies (\citealt{Battaglia_06}, Hendricks et al. in prep) shows distinct peaks in [Fe/H] around $-0.5$, $-1.0$, $-1.4$ and $-2.0$\,dex, which can be interpreted as an inhomogeneous, bursty SFH. In contrast, photometric SFH studies find continuous star formation (SF) from a population older than 10--12\,Gyr continuing to a young population not older than several hundred Myr with a significant SF peak around 3--4\,Gyr ago (\citealt{de_Boer_12}, \citealt{Coleman_08}). The age resolution of photometric studies, however, is limited due to the low color-sensitivity of old RGB stars and additionally biased by systematic uncertainties in the Calcium Triplet (CaT) metallicites used to break the age-metallicity degeneracy (\citealt{Battaglia_08}).

Possible proposed origins for the complex chemical evolution of Fornax are periodic interactions of the galaxy with the MW's tidal field (e.g., \citealt{Nichols_12}, see also \citealt{Pasetto_11} for Carina), or merger scenarios with other dwarf galaxies (\citealt{Coleman_04}, \citealt{Yozin_12}).
On the other hand, potential SF bursts may also be a consequence of the reaccretion of gas previously blown out of the galaxy by stellar winds with a time delay of several $10^9$\,years (e.g., \citealt{Revaz_09}, \citealt{Ruiz_13}). Interestingly, two of the observed peaks in the field star MDF roughly coincide with the metallicity of Fornax' GCs and it has been proposed that these peaks are not caused by intrinsic star formation variations, but rather by dissolved GC stars (\citealt{Larsen_12}). 
These arguments show that the SFH of Fornax is not well understood yet, and an important question remaining is whether the system evolved in unperturbed isolation, or whether its field population is the result of merging events and/or tidal interactions with the MW.

The $\mathrm{\alpha}$-elements (e.g., Mg, Si, Ca, Ti) provide a crucial tool for understanding a galaxy's chemical enrichment history. They are synthesized mainly in massive stars ($M_{\star} \geq 8M_{\sun}$) with a lifetime of not more than 30\,Myr, and distributed throughout a galaxy during the subsequent core-collapse SN\,II explosion. In contrast, SNe Ia are thought to produce primarily Fe--peak elements and may not contribute to a system's interstellar medium (ISM) for $\sim10^{9}$ years. Therefore, the $\mathrm{[\alpha/Fe]}$-ratio is a tracer for the relative contribution of SNe\,II and SNe\,Ia to the ISM. At early times, only SN\,II enrich the ISM which is observed as an $\mathrm{\alpha}$-rich plateau at very low values of [Fe/H]. The sudden drop in the $\mathrm{[\alpha/Fe]}$-ratio caused by the onset of SN\,Ia is generally called the ``knee" in the evolution of $\mathrm{\alpha}$-elements. Consequently, the knee in a plot of $\mathrm{[\alpha/Fe]}$ versus [Fe/H] is an indicator for the system's efficiency to build up and retain metals within its potential well (e.g., \citealt{Tinsley_79}, \citealt{Matteucci_90}, \citealt{Venn_04}, but see \citealt{McWilliam_13} for a different view).

It has been shown for several dSphs, that the $\mathrm{\alpha}$-element evolution is significantly different from the MW. Generally, their knee---if detected at all---lies at lower [Fe/H] than for MW halo field stars (see, e.g., \citealt{Tolstoy_09}), and the individual $\mathrm{\alpha}$-element ratios become more depleted (e.g., \citealt{Letarte_10}, \citealt{Koch_08}, \citealt{Sbordone_07}). Additionally, the position of the knee varies between individual dSphs (e.g., \citealt{Cohen_09, Cohen_10}). This apparent variation in the chemical enrichment process has been linked mainly with the total stellar mass of the individual galaxy, where faint, low-mass dwarfs show lower enrichment efficiencies. The same simplified argument also applies for the observed difference between dwarfs and the MW (\citealt{Matteucci_90}).

Unfortunately, for most of these galaxies, the detailed evolution of the $\mathrm{\alpha}$-elements is not well known yet. Until today the observed sample of metal-poor (MP) stars in dSphs with high-resolution (HR) spectroscopy is small. 
One reason for the lack of data in this metallicity regime is the generally small fraction of MP stars ($\mathrm{[Fe/H]}\leq-2.0$) in dSphs (\citealt{Helmi_06}). Second, to optimize galaxy membership, most large surveys target their central regions, which are known to be more metal rich than the outer parts (e.g., \citealt{Battaglia_06}, \citealt{Koch_06}, \citealt{Kirby_11c}). Although the few published abundances suggest that, at the metal-poor end, the $\mathrm{\alpha}$-elements in dSphs overlap with the metal-poor MW halo (\citealt{Shetrone_03}, \citealt{Tafelmeyer_10}), the only galaxy for which the position of the knee is well-defined so far is Sculptor (\citealt{Starkenburg_13}). For Fornax, \citet{Letarte_10} provides the only existing HR study. They determined the $\mathrm{\alpha}$-depleted level for metal-rich stars, but only detected very few stars with $\mathrm{[Fe/H]}\leq-1.3$, which did not allow for the determination of the knee, nor the chemical evolution at lower metallicities.

Here, we show the results for three $\mathrm{\alpha}$-elements (Mg, Si, Ti) in Fornax, determined from HR spectroscopy, covering the full range in [Fe/H]. We are able to define the position of the knee and trace the evolution of the $\mathrm{\alpha}$-elements from the $\mathrm{\alpha}$-rich, metal-poor plateau, to the $\mathrm{\alpha}$-depleted level at the metal-rich side of the knee. 
In Chapter \ref{chap_02}, we summarize our data and the reduction steps to obtain our abundances, which are subsequently presented in Chapter \ref{chap_03}. In Chapter \ref{chap_4}, we compare our results to chemical evolution model predictions. Finally, in Chapter \ref{chap_05} we will discuss our findings in regard to possible formation and evolution scenarios of Fornax.

\section{Data}
\label{chap_02}
\begin{figure}[h]
\begin{center}
\includegraphics[width=0.5\textwidth]{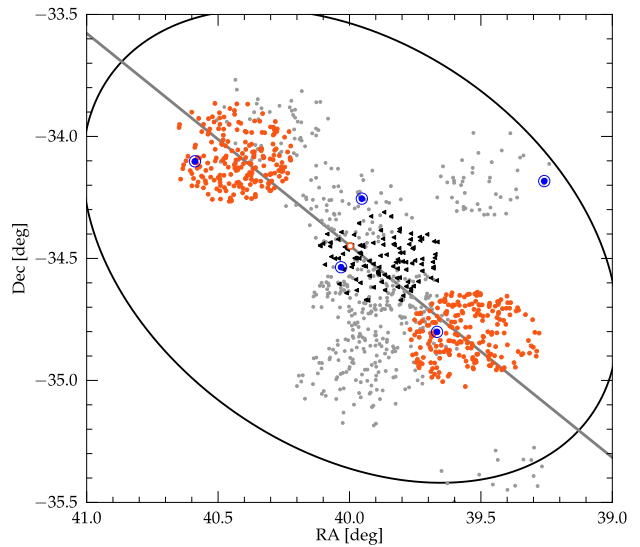}
\caption{Location of our targets (red dots) in the field of Fornax. GCs are marked as blue symbols, and previous spectroscopic studies using low-resolution (\citealt{Battaglia_06}; gray dots) and high-resolution (\citealt{Letarte_10}; black triangles) are also shown for comparison. To guide the eye, the nominal tidal radius with $r_{t}\approx1.15\degree$ is shown.}
\label{fig_1}
\end{center}
\end{figure}

Our sample of 431 targets in Fornax were selected from optical $V$ and $I$ broadband photometry (\citealt{Walker_06}) within a broad selection box around the red giant branch, spanning down to the horizontal branch magnitude. The targets are distributed in two opposite fields along the major axis of the galaxy, aiming specifically for stars in the outer part of Fornax which has a higher fraction of ancient, metal-poor stars, compared to the central region (\citealt{Battaglia_06}). Our fields also cover two of the five known GCs of Fornax (H2 and H5, \citealt{Hodge_61}). Figure\,\ref{fig_1} shows the location of our targets in comparison to previous, comprehensive high- and low-resolution studies. The spectra have been obtained in November 2008 with FLAMES at the VLT, where we used GIRAFFE in MEDUSA high-resolution mode (HR 21, $R\sim 16,000$, $8484 - 9001$\,\AA). With a total integration time for each pointing of 8 hours we obtain a typical signal-to-noise ratio (S/N) of $20-50$ per pixel. 

To extract and calibrate our spectra, we use \emph{girBLDRS} (GIRAFFE Base-Line Data Reduction Software, Geneva Observatory; \citealt{Blecha_03})\footnote{http://girbldrs.sourceforge.net}. The individual reduction steps include flat-field correction, bias subtraction and dark corrections, as well as wavelength calibration. Each exposure has been split up in 8 frames and we median-combine them in order to remove cosmic rays and other artifacts in the spectra. Since the near-infrared is strongly affected by sky emission features, a proper subtraction is crucial for accurate spectral analysis. For this reason, we have written a code which accounts for the different flux throughput in each particular fibre with respect to the dedicated skyfibres and also for small wavelength shifts between the sky- and the target fibres by matching only the strongest few emission features via $\chi^2$-minimization and subsequently apply the same scaling to the whole spectrum.

For the majority of our spectra, we were able to derive metallicities from the Calcium Triplet ($\lambda_{CaT_1}=8498.03\AA$, $\lambda_{CaT_2}=8542.09\AA$, $\lambda_{CaT_2}=8662.14\AA$). We use the sum of a Gaussian and a Lorentzian function to fit the profile of the second and third CaT lines and determine equivalent widths (see, \citealt{Cole_04}, or \citealt{Koch_06}). 
Classically, the calibration from CaT equivalent width to [Fe/H] is based on GC measurements (e.g., \citealt{Armandroff_88}, \citealt{Armandroff_91}, \citealt{Rutledge_97}). It has been shown in several studies, that these GC-based calibrations of the CaT metallicity is in good agreement with high-resolution abundance measurements from iron lines only in the metallicity regime between $-2.0\leq \mathrm{[Fe/H]}\leq -0.5$ (\citealt{Battaglia_08}, see also Hendricks et al. (in prep.)).
To obtain [Fe/H] from the CaT EWs, we therefore use the recently published calibration-equations by \citet{Carrera_13}, who made a dedicated effort to extend the calibration range to metallicities as low as $-4.0$\,dex.

We determine radial velocities for each star by comparison to a synthetic CaT template spectrum using \textit{iraf.FXCOR}, which yields a precision of $\sim$2--3 km s$^{\rm -1}$. The derived RVs are then used to weed out foreground stars and background galaxies, with a similar iterative clipping procedure as described in \citet{Walker_06}.

Since two Fornax GCs were included in our target fields, and the chemical enrichment history of GCs can be significantly different from that of the host galaxy, we exclude from our present sample stars within $60\arcsec$ (equivalent to $\sim4$ cluster core-radii) of the cluster centers.

\subsection{The alpha elements}

Individual abundances are determined using \emph{SPACE} (Stellar Parameters and Chemical abundances Estimator, \citealt{Boeche_13}, Boeche et al. in prep). This new code is the evolution of the \emph{RAVE} chemical pipeline (\citealt{Boeche_11}, \citealt{Kordopatis_13}), capable of deriving stellar parameters and elemental chemical abundances during the same analysis process. It uses a library of Generalized Curve Of Growths which are the extension of the well known curve of growths in the 3-dimensional stellar parameter space with variables $\mathrm{T_{eff}}$, log\,g, and [X/H].
Because the relatively low S/N and the limited wavelength range of our spectra does not allow a robust estimation for the atmospheric parameters within the code, we derive these parameters from optical $V-I$ colors, using the empirical calibration equations given in \citet{Alonso_99} with a reddening law of $A(V)/E(B-V)=3.1$ and a line-of sight reddening $E(V-I)=0.04$ (\citealt{McConnachie_12}). 
Micro-turbulence is assumed to be a function of $\mathrm{T_{eff}}$, and log\,g, and is calculated using a third-order polynomial given in \citet{Boeche_11}.

\subsection{Statistic and systematic uncertainties in chemical abundances}
The major source for systematic uncertainties in the derived abundances comes from the photometric estimation of $\mathrm{T_{eff}}$ and log\,g. To estimate this effect, we propagate the uncertainty of our photometry to $\mathrm{T_{eff}}$ and log\,g and rerun the abundance code on our full sample by separately varying each parameter. By that, we obtain individual systematic uncertainties for each star, defined by its specific photometric error. Typical values for $\delta\mathrm{[Fe/H]}$ and $\delta\mathrm{[X/Fe]}$ are below 0.1\,dex for all abundance ratios.
 
We estimate the statistical uncertainty for individual abundances with a subset of the synthetic spectra compiled by \citet{Kirby_11}.
These spectra cover a broad range in metallicities and additionally offer the option to vary the $\mathrm{\alpha}$-abundance between $\mathrm{[\alpha/Fe]}=-0.2$ and $+0.5$\,dex. First we trim the spectra to our observed wavelength range and convolve them to the resolution and pixel scale of our data. Then we create a set of 50 spectra for each point in the parameter space and add random poissonian noise mimicking a S/N of 30, a typical value for our observed spectra. The standard deviation in the derived abundance for each set eventually serve as our estimate for the random error.
Note, that we used spectra with $\mathrm{[\alpha/Fe]}=0.0$, but the dependance of the error on the actual $\mathrm{\alpha}$-abundance is small.
Typically, we find the combined uncertainty from all discussed aspects for our abundances to be smaller than 0.15\,dex in the metal-poor regime, and smaller than 0.1\,dex in the metal-rich regime (see Figure\,\ref{fig_2}).

\section{Results}
\label{chap_03}
\begin{figure*}[PH]
\begin{center}
\includegraphics[width=0.9\textwidth]{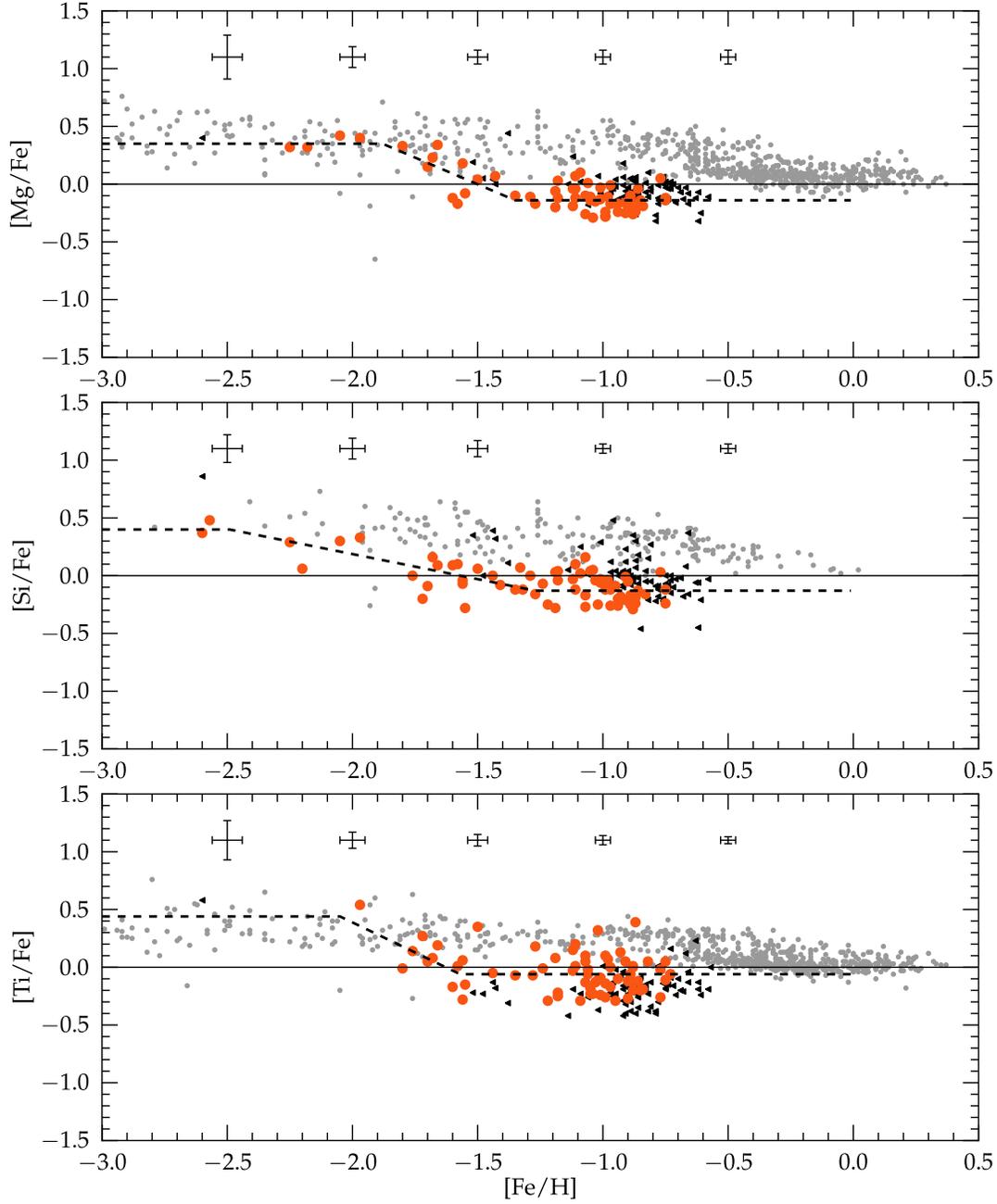}
\caption{Evolution of three $\mathrm{\alpha}$-elements in the Fornax dSph (from top to bottom; Mg, Si, Ti). Our sample from two outer fields is shown with red symbols and is supplemented with previous results for inner-field stars (\citealt{Letarte_10}; black triangles). Additionally, the evolution of $\mathrm{\alpha}$-elements in the MW halo and disks (\citealt{Venn_04}) is shown. Typical statistical uncertainties for our abundances are shown in the top of each panel. Systematic errors caused by photometric estimation of atmospheric parameters are typically smaller ($\leq0.1$\,dex). A toy model (see text) is overplotted for each element and shows a clearly defined knee in the evolution of Mg at $\mathrm{[Fe/H]}\approx -1.9$. The position of the knee is less clearly defined for Si and Ti, but can be ruled out to be more metal rich than -1.8\,dex.}
\label{fig_2}
\end{center}
\end{figure*}

The derived abundance ratios for three $\mathrm{\alpha}$-elements Mg, Si, and Ti are shown in Figure\,\ref{fig_2}. Here, we only use stars with a $\mathrm{S/N}\geq25$ and also removed results, for which our code only converged with $\chi^2$-values greater than three sigma from the mean. From the described selection, we obtain a sample of 58, 69, and 67 stars with measured Mg, Si, and Ti, respectively.

From our data we sample the $\mathrm{\alpha}$-distribution continuously between $\mathrm{[Fe/H]}\geq-2.5$ and $\mathrm{[Fe/H]}\leq-0.7$\,dex. As can be seen in Figure\,\ref{fig_2}, a knee in the distribution around $\mathrm{[Fe/H]}\approx~-1.9$ is clearly visible, especially in the evolution of Mg. For the other two elements the sampling in the metal poor area is less clear, but it is evident that both Si and Ti are already depleted to a sub-solar level ($\mathrm{[\alpha/Fe]}\leq0.0$) at metallicities below -1.5\,dex. Therefore, from Si and Ti it is still possible to set a strong upper limit for the iron abundance that marks the onset of SN\,Ia. 
We also find stars more metal poor than the knee to lie on the same $\mathrm{\alpha}$-rich plateau observed in the MW halo, whereas stars above $\mathrm{[Fe/H]}\approx-1.4$ seem to lie on a depleted plateau, significantly below the MW level at a corresponding iron abundance. Note that our mean abundances agree very well with the values derived by \citet{Letarte_10} at the overlapping metal-rich end. Since their sample has been taken exclusively from the central part of the galaxy, while our data come from outer fields, the agreement between the two datasets indicates that there is no difference in the level of depletion as a function of galactocentric distance or stellar density, at least within 2-3 half-light radii of the galaxy.

Although Mg, Si, and Ti are expected to share a common general evolution, their exact origin differs; while Mg is almost exclusively synthesized in SNe II, Si and Ti may also be produced in SN\,Ia explosions. Therefore, we might expect the decline in [Mg/Fe] after the knee to be steeper than [Si/Fe] and [Ti/Fe] (see, e.g., \citealt{Lanfranchi_04}). In our sample, we observe this difference most clearly between Mg and Si (see also Table\,\ref{tab_2}). 
If the position of the knee, however, solely depends on the SNe Ia time-delay, its location should be the same for all stars sharing the same chemical enrichment environment.

In order to quantify our results, we first construct a toy model (see \citealt{Cohen_09}), which uses the evolution of $\mathrm{\alpha}$-elements in the MW halo as template and assumes a constant plateau for stars more metal poor than the knee, and a second plateau for $\mathrm{\alpha}$-depleted stars. The two plateaus are linked with a linear slope. Accordingly, we create a piecewise function with the level of both plateaus in $\mathrm{[X/Fe]}$ ($P1$, $P2$), the onset of the knee ($K$) and the slope of the linear decline ($s$) as free parameters, which we subsequently derive by error-weighted $\chi^2$-minimization. The result is overplotted on the abundance pattern for each Element in Figure\,\ref{fig_2} and the corresponding parameters are listed in Table\,\ref{tab_2}.

\begin{table}[h]
\centering
\caption{Fiducial points for the chemical evolution of Mg, Si, and Ti in Fornax from our toy model.}
\begin{tabular}{ccccc}
\hline
\hline

Element					&		$K$					&		$ P1$ 		&  $P2$					&	$s$					\\\hline
[Mg/Fe]					&		$-1.88$				&		$0.35$		&	$-0.14$				&		$-0.93$			\\\
[Si/Fe]					&		$-2.49$				&		$0.40$		&	$-0.13$				&		$-0.43$			\\\
[Ti/Fe]					&		$-2.05$				&		$0.44$		&	$-0.06$				&		$-1.04$			\\\
[$\mathrm{\alpha}$/Fe]		&		$-2.08$				&		$0.35$		&	$-0.13$				&		$-0.67$			\\\hline

\end{tabular}
\tablecomments{For Si and Ti, no clear plateau is visible until the lower end of our sampled metallicity range, and the derived parameters are ambiguous. To derive the fiducial points for $\mathrm{[\alpha/Fe]}$ we combine all three elements where possible, and otherwise use any combination of only two species.}
\label{tab_2}    
\end{table}

\begin{figure}[h]
\begin{center}
\includegraphics[width=0.5\textwidth]{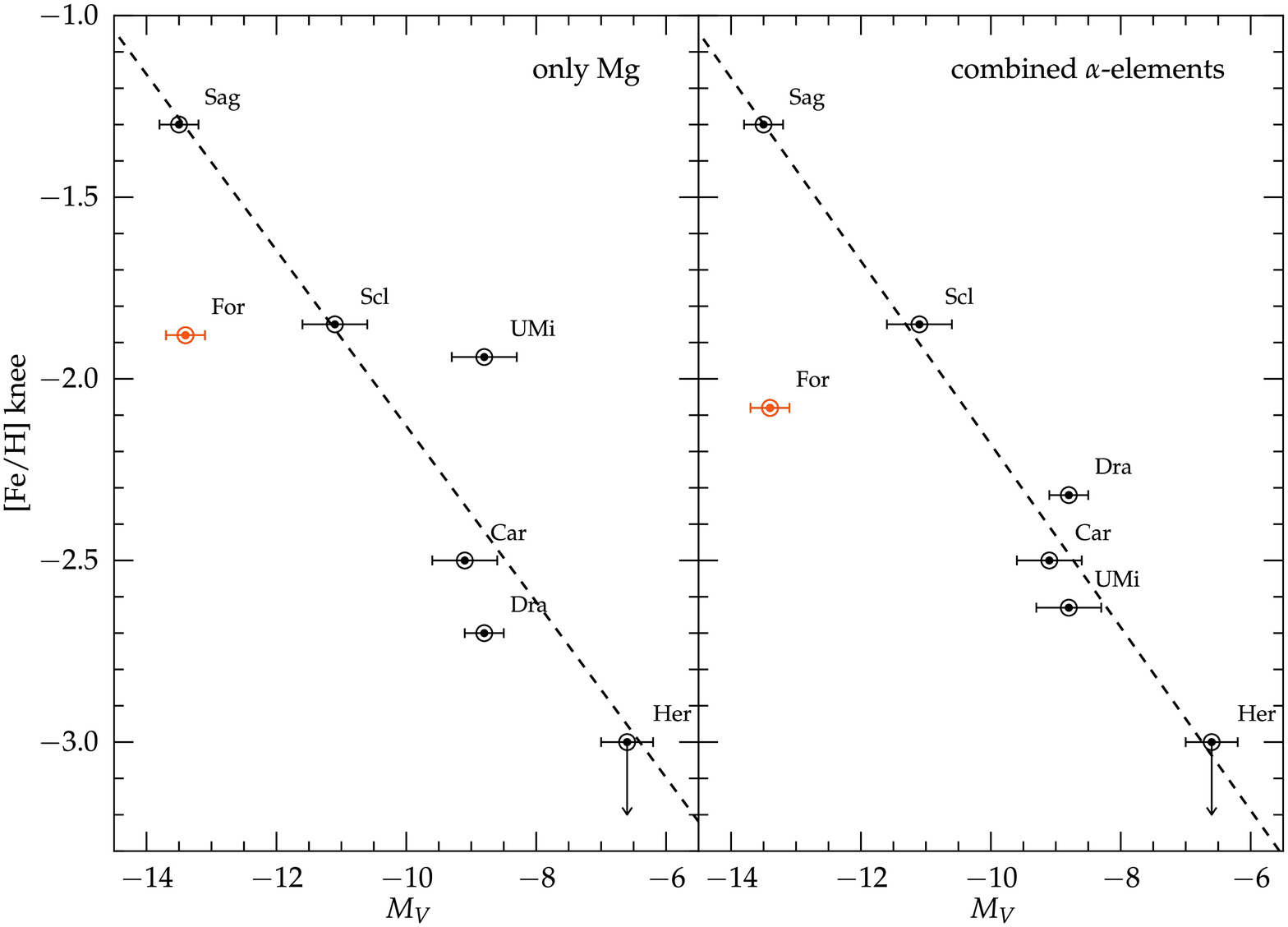}
\caption{Position of the knee in the $\mathrm{\alpha}$-element distribution in several dwarf galaxies as a function of absolute magnitude. Uncertainties in $M_V$ have been adopted from \citet{McConnachie_12}. Note, that Sagittarius might have been as much as two magnitudes more luminous in the past (\citealt{Niederste-Ostholt_10}). In the left panel we only use [Mg/Fe], in the right panel we use a combination of all available $\mathrm{\alpha}$-elements. 
The dashed line indicates the best fitting linear relation, when we exclude Fornax from the sample. The metal-poor knee of Fornax does not fall in an otherwise linear relation and either questions the formation scenario of this galaxy, or the understanding of chemical enrichment in dSphs.}
\label{fig_3}
\end{center}
\end{figure}

Most remarkably, the onset of the knee is located at $K=-1.88$\,dex for Mg, and $-2.08$\,dex for a combination of all available elements. 
The very low [Fe/H] at which we observe the knee in Fornax is unexpected for two reasons. First, chemical evolution models in previous studies that reproduce the observed MDF, consistently predict a knee at $\mathrm{[Fe/H]}\approx-1.4$, 0.5\,dex higher than what we find from our data (\citealt{Kirby_11a}, Lanfranchi et al. submitted). The second puzzling point arises when we compare our results to the findings from other dSphs, and specifically to Sculptor. For the latter galaxy the position of the knee is well determined at $\mathrm{[Fe/H]}\approx-1.8$ (\citealt{Starkenburg_13})---similar to what we find for Fornax--- although Sculptor's stellar mass is estimated to be around 10 times smaller. Vice versa, Sculptor's mean metallicity is lower by more than 0.6\,dex, although the evolution of  $\mathrm{\alpha}$-elements indicates a similar enrichment efficiency.
Since we compare absolute magnitudes, and consequently luminous masses, it is important to note that recent estimates have attributed Fornax also a larger M/L-ratio than Sculptor (\citealt{McConnachie_12}).
This is in conflict with the concept of more massive galaxies being more efficient in building up heavy elements, with a knee consequently at higher [Fe/H]. Using absolute magnitudes from \citet{McConnachie_12} as well as previous estimations of knee-positions in other dwarf galaxies for Sculptor (\citealt{Starkenburg_13}), Draco, Ursa Minor, Carina, Sagittarius (\citealt{Cohen_09}, \citealt{Cohen_10}), and Hercules (\citealt{Vargas_13}), Fornax' knee clearly falls out of an otherwise fairly linear relation (see Figure\,\ref{fig_3}). If the same mechanism was at play in all the dwarfs, Fornax' stellar mass should not exceed a few $10^6$ solar masses. 
Since not all studies we selected here do define the knee with the same method we applied for Fornax, we do not add individual uncertainties for the position of the knee in [Fe/H], which are generally in the order of $\sim 0.2$\,dex for the brighter galaxies (Sagittarius, Fornax, Sculptor) and larger than that for the rest.
Note, that for the faintest dSphs like Bo\"otes or Hercules with $M_{V}\geq -7$, the expected drop in $\mathrm{[\alpha/Fe]}$ is at such low metallicities that a linear model without plateau and knee also gives a reasonable fit to the data (e.g., \citealt{Gilmore_13}, \citealt{Vargas_13}).

\section{Comparison to Chemical Evolution Models}
\label{chap_4}

Chemical evolution models are a useful tool to analyze the enrichment history of dwarf galaxies; by matching the predicted evolution to the observed chemical abundances it is possible to put constraints on different SF and enrichment scenarios.  
In the recent past, the chemical evolution of $\mathrm{\alpha}$-elements in dSphs have been modelled for many individual galaxies. In a series of papers Lanfranchi et al. (\citealt{Lanfranchi_03}, \citealt{Lanfranchi_04}, \citealt{Lanfranchi_10}) as well as \citet{Kirby_09} and \citet{Kirby_11a} consistently found that, in order to reproduce the observed abundances and the metallicity distribution in the galaxies, the models have to be characterized by a very low star formation efficiency compared to the MW Halo and solar neighbourhood to reproduce the low values of $\mathrm{[\alpha/Fe]}$, in combination with a strong and efficient galactic wind to explain the observed MDF and the lack of gas in these systems today.
For most systems (except Carina) a single SF period with a spatially and timely invariant SF efficiency has been used, yielding a good fit to the observed properties (including $\mathrm{[\alpha/Fe]}$, $\mathrm{[s,r/Fe]}$, the MDF, and the present day gas- and stellar mass).

Since our data provide a continuous sequence of abundances from $\mathrm{[Fe/H]\approx -2.5}$ to $-0.7$\,dex, we can make use of the \citet{Lanfranchi_03} chemical enrichment models, which are specifically developed to reproduce the properties of dSphs. These models adopt up-to-date nucleosynthesic yields for intermediate-mass stars and SNe (Ia and II) as well as the effects of SNe and stellar winds on the energetics of the ISM. 
The main features as well as the theoretical prescriptions of the model are described in detail in \citet{Lanfranchi_03} and at this point we only summarize the key features and the specifications we made in order to adjust the model to Fornax.

In the model, Fornax is supposed to be formed by infall of pristine gas until a mass of $\sim 5 \times 10^8 M_{\odot}$ is accumulated inside a radius of 450\,pc. During and after the infall stars are formed according to a Salpeter initial mass function (IMF; \citealt{Salpeter_55}) and a pre-defined SFH. 
The models allow the outflow of gas through galactic winds and assume infall of primordial gas in the formation of the galaxy, but do not assume inflow from external gas or reaccretion of previously expelled material. 
The predicted position of the knee is sensitive to mainly two key parameters. First, the star formation efficiency ($\nu$), which scales the star formation rate (SFR) in the galaxy to the amount of available gas. Second, the wind efficiency ($\omega_{i}$; the index indicates a differential treatment of individual chemical species) determines the relation between the SFR and a galactic wind which removes thermally heated gas from the galaxy as soon as it is surpassing its gravitational potential (see \citealt{Lanfranchi_07} for a detailed discussion). Since a strong galactic wind causes the removal of potentially star forming gas, $\nu$ and $\omega_{i}$ both regulate the actual SFR, and by that the chemical enrichment efficiency of the galaxy.

In a recent paper, Lanfranchi et al.\,(2014, submitted) could not distinguish between a continuous SF and a SFH characterized by several distinct episodes of star forming activity. The limited data sample for the $\mathrm{\alpha}$-elements in their study with almost no stars below $\mathrm{[Fe/H]}\leq-1.3$\,dex allowed no conclusion. In the following we will therefore present and discuss two conceptually different scenarios: First, we adopt a continuous SFH and in a second model we mimic an interrupted, bursty SF with changing key parameters. As we will see below, the exact choice of SFH will have an important impact on the predicted evolutionary signatures.

\subsection{Metallicity Distribution Function}
In order to construct different SF scenarios, we use Fornax' MDF to constrain the possible parameter space of $\nu$ and $\omega_{i}$ for the galaxy, while we simultaneously force the models to reproduce the present luminosity, stellar mass, as well as the absence of gas or ongoing SF.

\begin{figure}[bh]
\begin{center}
\includegraphics[width=0.5\textwidth]{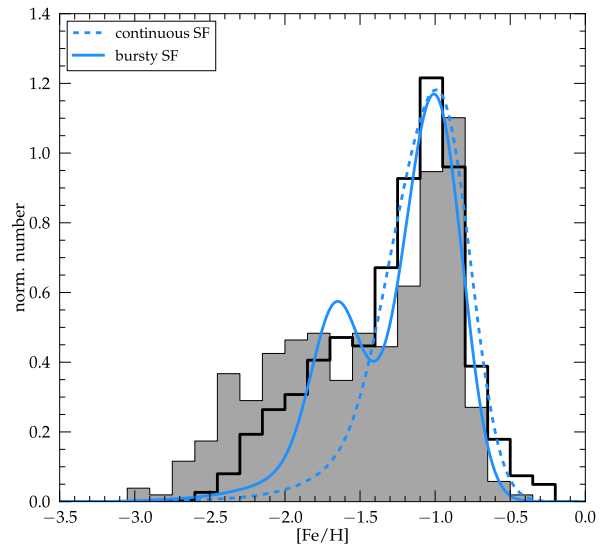}
\caption{Observed MDF in comparison to the chemical evolution model fits for a continuous (dashed blue line) and bursty SF (solid blue line). The filled histogram represents the local outer MDF, constructed from our sample is shown, while the outlined histogram uses the larger sample from \citealt{Battaglia_08} and is corrected for radial variations in the coverage (see main text). Both model fits are error-convolved by $0.15$\,dex in [FeH], a typical value for ours and Battaglia's CaT metallicities. While the observed distributions are normalized so that the integral over the area equals to one, we scale the models in order to give the lowest  $\chi^2$ to the global MDF.}
\label{fig_4}
\end{center}
\end{figure}

The MDF in Fornax shows a radial variation with a more metal-rich profile towards the center of the galaxy. Therefore the MDF derived from our outer sample is somewhat different compared to the one shown, e.g., in \citet{Battaglia_06}, and it is not trivial to decide which MDF reflects the chemical evolution of the $\mathrm{\alpha}$-elements. If we assume that the parameters for SF- and wind efficiency are mainly determined by \textit{global} properties like the Dark Matter (DM) halo mass, a global MDF would be the appropriate comparison. If instead these parameters are \textit{locally} defined for a specific radius or position in the galaxy, our abundances should be compared to the metallicity distribution from our sample. In Figure\,\ref{fig_4} we therefore show two versions of the MDF in Fornax: First, we derive a \textit{local} MDF from our sample, for which we use the CaT metallicities. Here, we apply the same selection criteria as for the $\mathrm{\alpha}$-elements (RV-membership, removal of possible GC stars), only with a lower threshold for the S/N ($\geq10$), yielding a sample of $\sim 350$ stars. The second version is a weighted MDF, for which we use the extended sample of CaT metallicities ($\sim1000$\,stars) from \citet{Battaglia_08} and use the same approach as outlined in \citet{Larsen_12} and \citet{Romano_13} to correct for the varying degree of completeness as a function of galactocentric distance. 
Note that the derived CaT metallicities in the two distributions shown here have been calibrated with different equations, and therefore small variations can be expected in the zero point and in the scaling of the MDF.

\begin{figure}[bh]
\begin{center}
\includegraphics[width=0.5\textwidth]{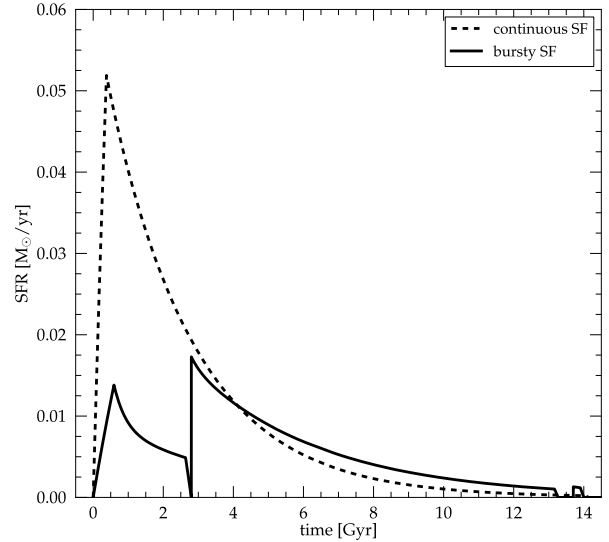}
\caption{Two different concepts for the SFH in Fornax: the dashed line shows the SFR as a function of time for our continuous model (see main text), and the solid line indicates the SFR resulting from the bursty model. The time is measured since the Big Bang.}
\label{fig_6}
\end{center}
\end{figure}

In each model fit, we searched for the parameters which best reproduce the peak at $\mathrm{[Fe/H]}\approx~-1.0$\,dex, which is the prominent feature in \emph{both} MDFs. Therefore, we consider the choice of $\omega_{i}$ and $\nu$ to be fairly robust against possible spatial biases in the MDF caused by incomplete sampling. The best fitting parameters for each model are summarized in Table\,\ref{tab_3} and the detailed SFHs are illustrated in Figure\,\ref{fig_6}. The model predictions for the MDF and the $\mathrm{\alpha}$-element evolution are shown in Figure\,\ref{fig_4} and \ref{fig_5}, respectively.

\begin{table*}[t]
\centering
\caption{Best fitting parameters for the chemical evolution models of the Fornax dSph}
\begin{tabular}{ccccccccc}
\hline
\hline

Model			&		Episodes of SF			&		  Periods (Gyrs)& 	$\nu \ (Gyr^{-1})$				&		$\omega_{i}$	&$\tau$ (Gyrs)	&$M_{init} \ (M_{\odot})$	&	$M_{final} \ (M_{\odot})$ &IMF	\\\hline
continuous		&		$1$					&		$0.0-14.0$		&	$0.380$					&		$-5.40$		& $2.42$	&	$5.0 \times 10^8$	&	$3.8 \times 10^7$	& Salpeter			\\\
bursty			&		$3$	&		$0-2.6$; $2.8-13.2$; $13.7-14.0$		&	$0.095$; $0.348$; $0.469$	&	$-5.65$			& $3.75$	&	$5.0 \times 10^8$	&	$2.2 \times 10^7$	& Salpeter		\\\hline

\end{tabular}
\tablecomments{Parameters for the best fitting models: SF efficiency ($\nu$), wind efficiency ($\omega_{i}$), and infall timescale ($\tau$). Note, that the actual SF in the models does not continue to the present day, due to the removal of the gas by galactic winds. The indicated periods for the SF episodes only serve as the input corner points for the model.}
\label{tab_3}    
\end{table*}

\subsection{Continuous Star Formation}

In the model with a continuous SF, the $\mathrm{\alpha}$-enhanced plateau at low [Fe/H] is caused by a pollution of the ISM from only SN\,II at early times. The model shows a slight decrease in the $\mathrm{[\alpha/Fe]}$-ratio already before the knee, since the average SN\,II progenitor mass is decreasing steadily with time, and the ISM is already polluted by a few Ia supernovae.
The onset of a significant number of SN\,Ia is observed as a knee in the $\mathrm{\alpha}$-evolution which causes a steep drop in the $\mathrm{[\alpha/Fe]}$-ratio. In the continuous model, this ratio keeps dropping in the subsequent evolution, since the rate of SN\,II is practically constant, while the number of SN\,Ia keeps rising.
At late times, this effect is enhanced by the onset of the galactic wind. The removal of gas decreases the SFR and, consequently, the number of SN\,II, which leads to lower injection of O and Mg (and to some extent Ca and Si) in the ISM. Iron, on the other hand, still pollutes the medium even after the end---or during an interruption---of SF due to the longer lifetimes of SN\,Ia progenitor stars.

The continuous model is still in agreement with our observations for $\mathrm{[Fe/H]}\leq-1.9$, but it predicts a knee not before $\mathrm{[Fe/H]}~-1.3$, about 0.5\,dex higher than we observe in the data.
For Mg and Si we find that the models systematically over-predict $\mathrm{[\alpha/Fe]}$ at a $2\sigma$ level between $-1.6 \leq \mathrm{[Fe/H]}\leq-1.1$. Ti has a somewhat larger observational scatter, and the mismatch is less pronounced.

It is important to stress, that we tested a variety of different parameter combinations in order to shift the knee towards lower [Fe/H], which we find only to be possible with a significantly lower SFR (either due to a low SF efficiency or a strong galactic wind at early times). This, however, is inevitably in disagreement with the high-metallicity peak in the iron distribution, existent in both the local and the global MDF, which can only be reproduced with a high SF efficiency and a late galactic wind. In other words, the high-metallicity peak in the MDF in combination with the metal-poor knee in the evolution of the $\mathrm{\alpha}$-elements \textit{rule out} an evolutionary set-up with continuous SF and constant SF efficiency.

Generally, a model with continuous SF also fails to reproduce a second plateau after the knee which we observe at a sub-solar level in our data. A possible reason for this mismatch could be an incompleteness-bias of our sample towards more $\mathrm{\alpha}$-enhanced stars, especially because \emph{SPACE} is only capable to derive abundances for $\mathrm{[\alpha/Fe]}\geq-0.3$ (lower abundances fall outside of the interpolation grid and will be flagged). However, this scenario is unlikely, since \emph{SPACE} gives the highest level of completeness at high metallicities, and we are able to derive abundances for more than 75\% of all stars with $\mathrm{[Fe/H]}\geq-1.0$ from our sample.

\subsection{Bursty Star Formation}

A bursty SF---that is a SFH with periods of intense SF separated by less active phases or even a complete shut down---presents a completely different scenario to the evolution of individual elements and the resulting MDF.
Now, we do not only enable the SF to shut off and on several times along the evolution, but also allow for \textit{different} SF efficiencies during each burst.

With only three individual bursts, it is now possible to reproduce all crucial features in the $\mathrm{\alpha}$-evolution, as well as the general appearance of the MDF. Generally, a gradual increase of the SF efficiency for subsequent bursts, makes it possible to bring the low [Fe/H] of the $\mathrm{\alpha}$-elements in accordance with the metal-rich peak in the MDF.

\begin{figure*}[PH]
\begin{center}
\includegraphics[width=0.9\textwidth]{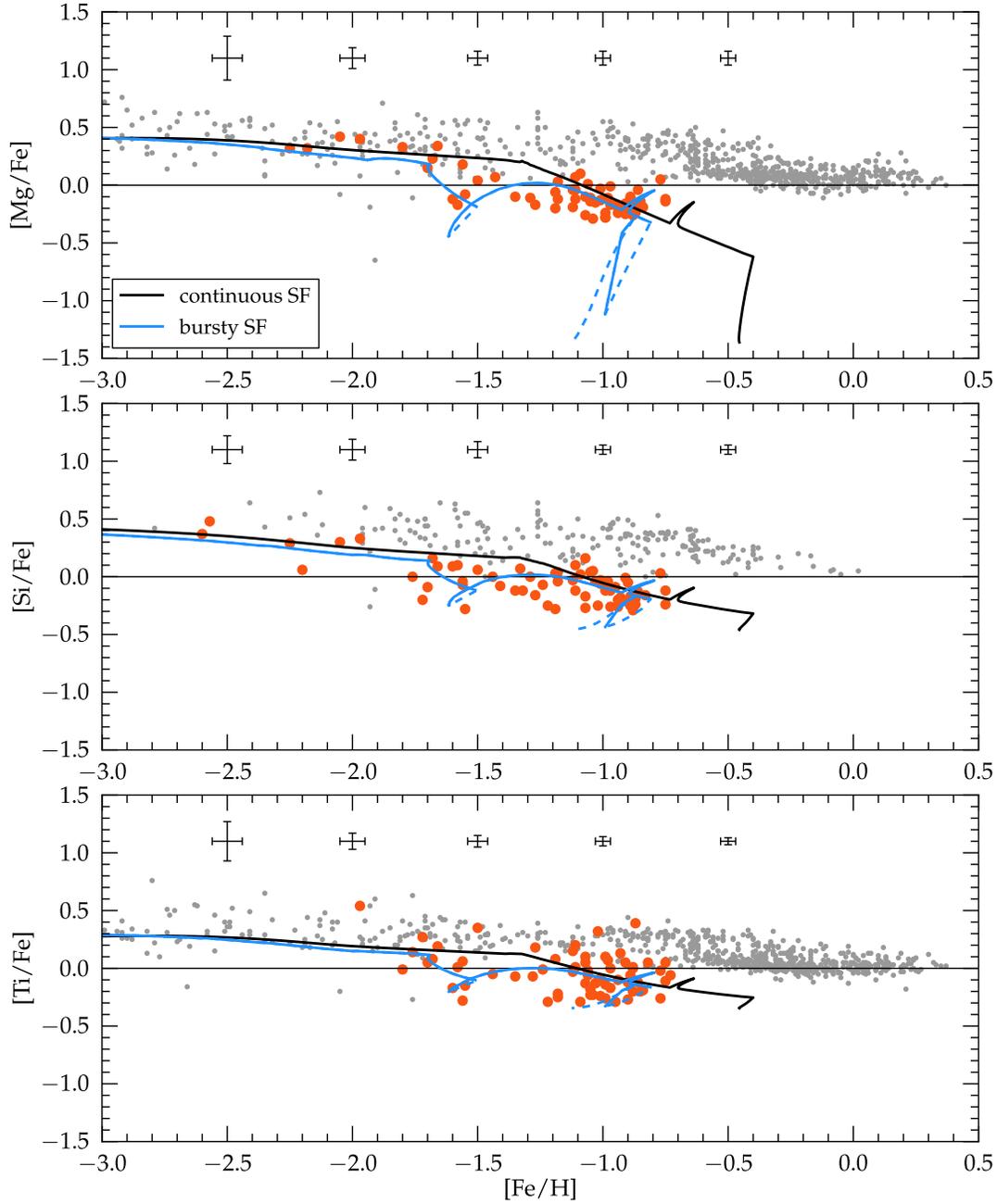}
\caption{
Model fits to the observed distribution of the $\mathrm{\alpha}$-elements Mg, Si, and Ti. The best fitting continuous model is shown in black, and the bursty SF model in blue. Times with no SF are indicated as dashed segments. While the continuous SF model predicts a knee in the evolution at [Fe/H] about 0.5\,dex too high, we can reproduce the metal-poor knee as well as the sub-solar plateau in the $\mathrm{\alpha}$-elements on the metal-rich side of the knee with a bursty model allowing for changes in the SF efficiency.}
\label{fig_5}
\end{center}
\end{figure*}

Similar to the continuous model, the plateau at very low metallicities is caused by the enrichment of the ISM exclusively from SNe\,II. But now, the knee occurs at significantly lower [Fe/H].
This is caused mainly by the low SF efficiency we assigned to the first SF episode, which prevents the galaxy to build up iron as fast as in the case of continuous SF. In addition, the pause of SF after 2.6\,Gyr is accompanied by a lack of young, massive SN\,II polluters at this point, which leads to an additional drop of the $\mathrm{[\alpha/Fe]}$-ratio in the ISM (see dashed evolution segments in Figure\,\ref{fig_5}). The onset of the second burst of SF 200\,Myr later with a higher SF efficiency has the opposite effect on the evolution of the $\mathrm{\alpha}$-elements; the average mass of a SN\,II explosion jumps up, which not only stops the abundance ratio to drop, but actually causes a temporary increase in $\mathrm{[\alpha/Fe]}$, observed as a bump in the evolution.
The third burst has a similar effect (visible at $\mathrm{[Fe/H]}\approx~-0.9$\,dex), but since it occurs at late times when the galactic wind has already removed the majority of the gas, we do not expect a large fraction of stars along this sequence.

The ``bouncing'-effect caused by the interruption in star formation is in very good agreement with our data. It is possible that the sub-solar plateau which we observe in the evolution of the $\mathrm{\alpha}$-elements---and which is also observed in other dSphs---is in fact the interpretation of one or several bumps, caused by interruption in the SF during the evolution of the galaxy, possibly accompanied by a change in the SF efficiency.

Note that, while both models do reproduce the metal-rich part (including the peak) in the MDF, the bursty SF scenario also yields the better fit to the observed distribution below $-1.5$\,dex (see Figure\,\ref{fig_4}). However, it still slightly underestimates the fraction of most metal-poor stars with $\mathrm{[Fe/H]}\leq-1.9$. The introduction of an additional, brief SF episode to the model at very early times with high SF efficiency would give a better fit to the data. Another possible explanation for the discrepancy between model and observed MDF at the lowest metallicities arises from the high specific frequency of GCs in Fornax: Recently, it has been proposed that Fornax' field population hosts a significant fraction of GC stars, stripped from the star cluster population associated with this galaxy (\citealt{Larsen_12}). These authors estimate that as much as 1/4 of the metal-poor population in this galaxy might originate from GCs. Four of the five existing globulars in Fornax have metallicities below -2.0 dex (e.g., \citealt{Strader_03}, \citealt{Larsen_13}, \citealt{Letarte_06}) and from the extended CaT metallicity sample provided in \citet{Battaglia_08}, we can estimate the relative stellar fraction with $\mathrm{[Fe/H]}\leq-2.0$ to be only $\sim0.08$.
A significant number of dissolved GC stars in the field could therefore plausibly explain the discrepancy between the observed and the model MDF.

\section{Discussion}
\label{chap_05}

We find a clear knee in the evolution of the $\mathrm{\alpha}$-elements as a function of [Fe/H]. Such a feature has been observed for field stars in the MW halo, and is also predicted by chemical evolution models as the delayed onset of SNe Ia.
The fact that we find Fornax' knee to be as metal-poor as the knee in the $\sim10$ times less massive and $\sim0.6$\,dex more metal-poor Sculptor dSph prompts the question, if, or to what extend the total stellar mass of these galaxies determined their general efficiency to build up heavy elements over time.

By comparing our results to chemical evolution models we find that a continuous SF cannot bring the knee in the $\mathrm{\alpha}$-elements into a consistent evolutionary scenario with the observed MDF in Fornax. In contrast, a bursty SFH with gaps between the individual SF episodes and significant variations between the SF efficiencies can not only explain the combination of the metal-poor knee and a metal-rich MDF, but at the same time provides an explanation for the $\mathrm{\alpha}$-depleted plateau in the data, which we find impossible to reproduce with a continuous SFH. 

Although it is surprising that a uniform SFH in Fornax does not agree with the observations, it is not striking that the match between model and data improves when the synthetic SFH is designed more flexible by allowing time-dependent key parameters. Therefore it is important to evaluate the assumptions we made for the model with respect to their physical implications for the galaxy and with regard to the the findings from previous studies.

The assumption of major SF episodes separated by several hundred\,Myr is not in agreement with the photometric SFH reconstruction in \citet{de_Boer_12}, who found a continuous SF present at all ages out to a radius of 0.8\,degree. On the other hand, MDFs constructed from large samples of CaT-measurements in \citet{Battaglia_06} and \citet{Coleman_08} show that the distribution of stars is not homogeneous, but peak at several---in fact three or four---metallicities. 
The MDF from our sample shows similar signatures at the same metallicities (see Figure\,\ref{fig_4})\footnote{A detailed population analysis, motivated by the findings presented in this work, will be part of a forthcoming study.}.
Additionally, \citet{Amorisco_12} find that each of the three more metal-rich populations can be distinguished by distinct kinematic properties, with a possible counter-rotation between the components, indicating a clear dynamical distinction between the populations.
Having in mind that the age resolution in photometric SFH studies gradually decreases (to several Gyr) towards older ages due to a smaller color-sensitivity of old RGB stars, together with the fact that iron abundances for stars from the CaT (as used in de Boer et al.) become systematically uncertain at lower metallicities, it is possible that an interruption of SF before the ISM is enriched to $\mathrm{[Fe/H]}\approx-1.5$ cannot be resolved in such studies.
Finally, several discrete SF episodes have been observed in the Carina dSph (e.g., \citealt{Monelli_03} an references therein), which gives an empirical validity for this evolutionary concept in dSphs.
Therefore, we consider the assumption of distinct SF bursts in Fornax to be reasonable, and the implementation of three episodes in good agreement with previous findings. 

The values for the best fitting SF and wind efficiency parameters are of comparable size to the corresponding best fitting parameters in other dSphs; generally the SF efficiency is low and the galactic wind intense compared to MW field stars. While the galactic wind for our model is on the lower end of what has been found for other dwarfs (there: $6.0 \leq \omega_{i} \leq 13.0$), the individual SF efficiencies for Fornax lie well in the middle of the typical parameter space in other studies, where best fitting values range between $\nu=0.03\,Gyr^{-1}$ for Draco and $\nu=3.0\,Gyr^{-1}$ for the massive Sagittarius dwarf. Sculptor, in comparison, is best fitted with $(\nu,\omega_{i})=(0.2,13.0)$.

However, the necessity of a strong variation of the SF efficiency (by a factor of $\sim5$) between the first and the third episode in order to explain the data is not a trivial assumption, especially since \textit{all} other dSphs (including Carina) could be successfully modelled with a uniform SF efficiency. What can cause such a drastic change in the SF efficiency in Fornax and the interruption between SF episodes, and which galactic parameters control its SFH?

A hint for a possible explanation comes from the radial metallicity gradient in Fornax (e.g., \citealt{Battaglia_06}). If the location of SF moves through the galaxy from outward in, and assuming that the SF efficiency depends on the radial position in the DM potential of the system, the net effect would be a time-dependence of the SF efficiency.
In such a scenario, the star formation could be altered and stopped periodically by radiative heating of the ISM from SNe\,II. Hydrodynamical simulations have shown, that SNe\,II explosions can transfer sufficient thermal energy into the medium to produce local cavities in the density distribution and halt the SF process until the gas has cooled and fallen back to reignite the SF process (e.g., \citealt{Ruiz_13}, \citealt{dErcole_99}, see also \citealt{Nichols_12} for an alternative explanation of periodic SF in dSphs).

Another evolutionary scenario is that Fornax experienced one, or several merger events with other gas-rich systems.
For massive galaxies it has been shown that the SFR can be enhanced by a factor of up to 10, when they interact in close pairs or mergers (\citealt{Scudder_12}). If this also holds for less massive systems, a merger scenario for Fornax
could lead to a variation in the SF environment with time, and simultaneously explain the gap in stellar mass between Fornax and Sculptor, despite their apparently very similar early chemical enrichment history.
In this case, the initial DM halo that defined the early chemical evolution of Fornax would have been less massive and might have formed initially only the metal-poor population of the MDF observed today. Through subsequent accretion events, the galaxy could have gained additional mass, and subsequently formed its enriched populations---with varying SF efficiencies. 

In fact, there are several studies supporting a merger event for Fornax, from a number of observational aspects.
\citet{Coleman_04} identify shell-like overdensities in their photometry and \citet{de_Boer_13} show that stars in these features are significantly younger than in its direct environment.
MDFs constructed in \citet{Battaglia_06} and \citet{Coleman_08} show, that the distribution of stars is not homogeneous, but peak at several metallicities. \citet{Battaglia_06} also find a distinct bimodality in the radial velocity distribution for the most metal-poor component, suggesting this subpopulation to be in an dynamically unrelaxed state. 
Note, that we find a similar distribution for this population in our kinematics (Hendricks et al. in prep).
The distinct kinematic properties for different populations in Fornax found by \citet{Amorisco_12} also support a merger scenario, and in fact these authors propose a ``bound-pair" as a likely evolutionary scenario.
In recent simulations dedicated to the complex kinematic and chemical structure observed in Fornax, \citet{Yozin_12} are able to reproduce both the photometric overdensities as well as the peculiar velocity bimodality for MP stars in a scenario where Fornax has experienced a merger event between 3.5 and 2.1\,Gyr ago.
Note, however, that a bimodal velocity distribution in old stars, as observed in Fornax, may also be the result of stripped GC stars in a triaxial DM potential profile (\citealt{Penarrubia_09}), whereas the fairly young shell structure ($\sim 2$\,Gyr) would require a significantly younger progenitor system.

Finally, \citet{Kirby_11c} found that amongst several simple analytical models, their MDF of Fornax and other dSphs is best reproduced when the model allows for the infall of external (pristine) gas onto the galaxy during its evolution.
The pure accretion of gas might have similar effects on the star forming environment in Fornax as mergers with a stellar component and therefore may also explain our findings.

The necessity of a bursty SFH with variations in the SF efficiency we propose here for the Fornax dSph arises from model predictions, and we do not know to what extent these models reflect the actual conditions in the galaxy. From observational side, the most promising way to test for the episodic nature of SF in Fornax---independent of its evolutionary interpretation---would be a deep photometric study of this galaxy, which is able to resolve the subgiant branch at $V\approx23.5$\,mag, and by that uncover possible distinct stellar population sequences. (e.g.,\citealt{Smecker-Hane_96}, or \citealt{Monelli_03} for Carina). Evidence for past mergers could be found in local overdensities of stars with the same velocity or [Fe/H] characteristic or galactic streams in the periphery of the galaxy. If the chemical enrichment efficiency inside Fornax changes with radius, we should expect to find different evolutionary sequences of, e.g., the $\mathrm{\alpha}$-elements from samples taken at different galactocentric distances. This, however, is a difficult endeavour since the fraction of metal-poor stars in the central region is extremely small. 

On the other hand, the chemical evolution in a merger scenario is associated with an inflow of stars as well as pristine or enriched gas to an existing system. It is therefore beyond the capabilities of the models we use here, and new chemical and hydrodynamic models are needed in order to support or rule out different evolutionary pathways. 

If Fornax actually grew from several smaller building blocks, it has to be asked whether this is an exception, or if it is a common evolutionary path for dwarf galaxies. In case Fornax instead did evolve in an isolated environment, and in fact did not gain a major part of its mass from external sources, the inevitable alternative is to introduce a new parameter to the chemical enrichment problem of this galaxy, such as a radial gradient of the SF and wind efficiency.

\begin{acknowledgements}
B. Hendricks thanks M. Frank and N. Kacharov for many fruitful discussions.
BH and AK acknowledge the German Research Foundation (DFG) for funding from Emmy-Noether grant Ko 4161/1. This work was in part supported by Sonderforschungsbereich SFB 881 "The Milky Way System" (subproject A5) of the DFG . G.A.L acknowledges brazilian agency CNPq, proj. 308677/2012-9. C.I.J. acknowledges support through the Clay Fellowship administered by the Smithsonian Astrophysical Observatory. This work was partly supported by the European Union FP7 programme through ERC grant number 320360.
\end{acknowledgements}

\end{document}